\documentclass[a4paper,fleqn,usenatbib]{mnras}
\usepackage[T1]{fontenc}
\usepackage{ae,aecompl}
\usepackage{graphicx}
\usepackage{amsmath}
\usepackage{amssymb}
\usepackage{xcolor}
\usepackage{xspace}
\usepackage{newtxtext,newtxmath}

\newcommand{\muc}{\mu_{\rm c}}
\newcommand{\muct}{\tilde{\mu}_{\rm c}}
\newcommand{\betac}{\beta_{\rm c}}
\newcommand{\gammac}{\gamma_{\rm c}}
\newcommand{\muo}{\mu_{\rm o}}
\newcommand{\betao}{\beta_{\rm o}}
\newcommand{\gammao}{\gamma_{\rm o}}
\newcommand{\cmb}{{\rm CMB}}
\newcommand{\rR}{{\rm R}}
\newcommand{\thetae}{\theta_{\rm e}}

\newcommand{\expf}[1]{{\rm e}^{#1}}

\title[Radio SZ modelling]{Refined modelling of the radio SZ signal: kinematic terms, relativistic temperature corrections and anisotropies in the radio background}

\author[E. Lee, J. Chluba \& G.P. Holder]{
Elizabeth Lee$^{1}$\thanks{E-mail: elizabeth.lee-2@postgrad.manchester.ac.uk},
Jens Chluba$^{1}$
and
Gilbert P. Holder$^{2,3}$
\\
$^{1}$Jodrell Bank Centre for Astrophysics, Department of Physics and Astronomy, The University of Manchester, Manchester M13 9PL, UK
\\
$^{2}$Astronomy Department, University of Illinois at Urbana-Champaign, 1002 W. Green Street, Urbana, IL 61801, USA
\\
$^{3}$Department of Physics, University of Illinois Urbana-Champaign, 1110 W. Green Street, Urbana, IL 61801, USA
}

\date{\vspace{-1mm}Accepted XXX. Received YYY; in original form ZZZ}

\pubyear{2021}

\begin{document}
\label{firstpage}
\pagerange{\pageref{firstpage}--\pageref{lastpage}}
\maketitle

\begin{abstract}
A significant cosmological radio background will inevitably lead to a radio Sunyaev-Zeldovich (SZ) effect. In the simplest limit, the combined signal from the scattered radio and cosmic microwave background exhibits a null at around $\nu \simeq 735$ MHz. 
Here, we show that kinematic and relativistic temperature corrections to this radio SZ signal are easily calculable. 
We treat both the cluster and observer motion, and the scattering of anisotropies in the radio background, highlighting how the spectrum of the radio SZ effect is affected in each case.
Although relativistic temperature corrections only enter at the level of a few percent, our expressions allow high-precision modelling of these terms.
By measuring the SZ signal around the radio null, one is in principle able to place constraints on the properties of a cosmological radio background. 
A combination with standard SZ measurements from large cluster samples could provide a promising avenue towards breaking degeneracies between different contributions. Stacking analyses can reduce the effect of kinematic corrections and dipolar anisotropies in the radio background, thereby providing a way to constrain the redshift dependence of the average radio background.
Our qualitative discussion is meant to give an analytic understanding of the various effects and also motivate further studies with the aim to obtain quantitative forecasts of their observability. At this stage, a detection of the corrections seems rather futuristic, but the advent of large SZ and X-ray cluster samples could drastically improve our ability to disentangle various effects.
\end{abstract}

\begin{keywords}
cosmology: theory -- galaxies: clusters: general -- radio continuum: general -- cosmic background radiation
\end{keywords}

\section{Introduction}

The Sunyaev-Zeldovich (SZ) effect \citep{Sunyaev1980} can be understood as a distortion to the cosmic microwave background (CMB) caused by Comptonisation with moving electrons. These electrons are typically found in the largest structures of the Universe, in particular galaxy clusters, where their temperatures often reach $\gtrsim 10^7$ K (or $\gtrsim 1$~keV). This greatly exceeds the temperature of CMB photons ($\simeq 2.7$ K), and as such electrons up-scatter the CMB photons, causing a decrement of photons at `lower' frequencies, an increase of photons at `higher' frequencies, and a null compared to the CMB at $\simeq 217$ GHz. Overviews of the traditional SZ effect and its use in cosmology and astrophysics can be found in \citet{Carlstrom2002} and \citet{Mroczkowski2019}.

Galaxy clusters (or other giant bodies of hot free electrons) can accordingly be considered as {\it operators} on any background source of photons \citep[e.g.,][]{Sunyaev1971, Kamionkowski1997b, Cooray2006, Kholupenko2015, Grebenev2020}. And any such source, assuming the photons have energies that are much lower than those of the scattering electrons \citep[see][for a detailed review of the Compton process]{Sarkar2019} will all gain a similar form of distortion from passing through the cluster. These distortions can all be considered separately, and then combined to understand the total distortion to the background photon distribution.

With this in mind, \citet{Holder2021} showed that a basic analysis of the low-frequency radio background would reveal a distortion which, combined with the standard CMB SZ signal, would exhibit a null at around $\nu \simeq 735$ MHz if the background is cosmological. The radio background itself has been considered several times over recent years \citep{Fixsen2011,Seiffert2011,Singal2018,Dowell2018}; however, its origin is still not well understood \citep[e.g.,][]{Chang2018,  Mittal2021}. Using the radio SZ effect, the origin of the radio background could be studied using methods similar to those employed for the standard SZ effect, but at much lower frequencies  \citep{Holder2021}.

In this work, we consider how kinematic and relativistic corrections to this combined radio and CMB SZ signal will appear at low frequencies. Effects from the scattering of radio background anisotropies are also discussed.
We show that a simple asymptotic expansion of the Compton collision terms is largely sufficient for all these calculations (see Sect.~\ref{sec:maths}). 
We explore how with sufficient knowledge of a cluster (which could in principle be measured from the traditional CMB SZ signal) we can use measurements of the radio SZ signal to determine details of the radio background itself.

Much of the physical intuition is completely analogous to the standard SZ effect; however, due to the steepness of the radio background spectrum several interesting differences arise as we highlight here.
Our simple expressions allow us to include all of the effects with just a few parameters.
This opens a way to high-precision modelling using future large SZ cluster samples.

\section{The radio SZ signal}
\label{sec:maths}
In this section, we derive expressions for the radio SZ signal that include kinematic and relativistic temperature corrections. We also show how anisotropies in the radio background affect the signal. The analysis is based on an asymptotic expansion of the Compton collision term. For the standard SZ effect, these expansions have been written in many forms \citep{Sazonov1998, Challinor1998, Itoh98}, but here we will follow the formulation of \citet{Chluba2012SZpack}, henceforth CNSN. 

\subsection{A summary of the SZ formalism}
The photon phase space density $n(t, \nu, \pmb{\hat{\gamma}})$, dependent on frequency $\nu$ in the direction $\pmb{\hat{\gamma}}$, varies over time according to the Boltzmann equation. Ignoring stimulated scattering terms, this allows us to write \citep[e.g.,][]{Pomraning1972, Buchler1976, Chluba2012SZpack}
\begin{equation}
    \frac{\partial n}{\partial t} \simeq c \int \frac{{\rm d}\sigma}{{\rm d} \Omega'} [f' n'-f n] {\rm d}^2 \hat{\gamma}' {\rm d}^3 p,
\end{equation}
where ${\rm d}^2 \hat{\gamma}'$ is the solid angle over incoming photons and $p$ the electron momentum. We have written $n = n(\nu, \pmb{\hat{\gamma}})$, $n' = n(\nu', \pmb{\hat{\gamma}}')$ for convenience, and similarly $f=f(\pmb{p})$,  $f'=f(\pmb{p}')$ where $f(\pmb{p})$ is the electron phase space distribution. 
We assume that the photon and electron distributions are isotropic. $\frac{{\rm d}\sigma}{{\rm d} \Omega'}$ is the differential scattering cross section for Compton scattering. In SZ studies, we ignore photon recoil and Klein Nishina corrections, valid because $h\nu\ll k_{\rm B} T_{\rm e}$.

Assuming that the electrons are fully thermalised, we can model their phase space density as a relativistic Maxwell-Boltmann distribution, which has the form
\begin{equation}
    f(\pmb{p}) = \frac{N_{\rm e} \expf{-\gamma/\theta_{\rm_e}}}{4\pi (m_{\rm e} c)^3 K_2(1/\thetae) \thetae}
\end{equation}
for total electron number density $N_{\rm e}=\int {\rm d}^3 p \,f(\pmb{p})$.
Here, $K_2$ is the modified Bessel function of the second kind, and the dimensionless temperature, $\thetae=k_{\rm B} T_{\rm e}/m_{\rm e} c^2$. In terms of the dimensionless momentum, $\eta=|\pmb{p}|/m_{\rm e} c$, we furthermore have $\gamma = \sqrt{1+\eta^2}$ for the Lorentz factor of the electron. 

Introducing the dimensionless frequency, $x=h\nu/k_{\rm B} T_\cmb$ and expressing $n(x',\pmb{\hat{\gamma}}')$ in terms of spherical harmonic functions $Y_{\ell m}(\pmb{\hat{\gamma}}')$ with coefficients $n_{\ell m}(x')$, we can rewrite the Boltzmann equation by expanding in terms of the frequency shift $\Delta_\nu =(\nu'-\nu)/\nu\ll 1$:
\begin{equation} 
\label{eq:Dn_complete}
\begin{split}
    &\frac{\partial n(x,\pmb{\hat{\gamma}})}{\partial \tau} \simeq -n(x,\pmb{\hat{\gamma}})+\sum_{k=0}^\infty \sum_{\ell,m} I^k_{\ell m} x^k \partial^k_x n_{\ell m},\\
    &I^k_{\ell m}=\frac{1}{N_{\rm e} \sigma_{\rm T}\,k!}\int \frac{{\rm d}\sigma}{{\rm d}\Omega'} f(\pmb{p}) \,\Delta_\nu^k \,Y_{\ell m}(\pmb{\hat{\gamma}}'){\rm d}^2 \hat{\gamma}' {\rm d}^3 p,
\end{split} 
\end{equation}
where $\tau=\int N_{\rm e} \sigma_{\rm T} c \,{\rm d}l$ is the Thomson optical depth along the line of sight.
The kernel moments $I^k_{\ell m}$ are only functions of the electron temperature (for $h\nu\ll k_{\rm B} T_{\rm e}$) and independent of the energy-dependence of the photon distribution, which is encoded by the terms $x^k \partial^k_x n_{\ell m}$.

\subsection{The asymptotic expansion}
\label{sec:asympt}
We can consider the moments $I^k_{\ell m}$ in terms of a temperature expansion about $\thetae\ll 1$.  Up to a fixed temperature, this truncates the expression in Eq.~\eqref{eq:Dn_complete}.
At high-frequencies ($x\gtrsim 5$), this approach does not converge for the standard SZ effect (see CNSN for in depth discussion);
however, at low frequencies (as required for the radio SZ effect), this expansion remains accurate to extremely high cluster temperatures (i.e., $T_{\rm e} \simeq 40$~keV or $\thetae \simeq 0.07$), as we illustrate here.

Assuming there is negligible {\it CMB frame} anisotropy in the unscattered photon distribution, i.e., $n(x, \pmb{\hat{\gamma}})\approx n(x)$, and ignoring effects of polarisation (since they will be small), all anisotropy in the system will accordingly come from the motion of the cluster relative to the CMB rest frame.
For a cluster moving at dimensionless speed $\betac = \varv_{\rm c}/c$ and in a direction $\muct=\cos \theta_{\rm c}$ relative to the line of sight as measured in the {\it cluster frame}, we can find that the change in the photon occupation number due to Compton scattering is given by [compare Eq.~(12) of CNSN] 
\begin{equation} \label{eqn:Dn_asym} \begin{split}
    \Delta \tilde{n} \simeq &\tau \thetae\sum_{k=0}^\infty \thetae^k \tilde{\mathcal{Y}}_k + \tau \betac^2 \thetae\sum_{k=0}^\infty \thetae^k \tilde{\mathcal{M}}_k \\
    &\quad +\tau \muct \betac \left(- \tilde{x}\partial_{\tilde{x}}\tilde{n}+\thetae\sum_{k=0}^\infty \thetae^k \tilde{\mathcal{D}}_k\right)\\
    &\qquad +\tau P_2(\muct)\betac^2\left(-\frac{3}{10} \tilde{x}^2\partial_{\tilde{x}}^2 \tilde{n} + \thetae \sum_{k=0}^\infty \thetae^k \tilde{\mathcal{Q}}_k\right).
\end{split} \end{equation}
Here, the Compton-$y$ parameter is, $y = \int \thetae N_{\rm e} \sigma_{\rm T} c {\rm d}t = \tau \thetae$. We also used the Legendre polynomial $P_2(\mu) = 3(\mu^2-1)/2$. In Eq.~\eqref{eqn:Dn_asym}, tildes [e.g., $\muct$, $\tilde{n}=n(\tilde{x})$] indicate that these quantities are being calculated in the cluster frame. We dropped this distinction for the electron temperature and number density that enter the scattering $y$-parameter, but emphasize that these are cluster frame quantities (see CNSN for in depth discussion).

In practice, the sums in Eq.~\eqref{eqn:Dn_asym} converge quickly at $\tilde{x}\ll 1$, and summing up to $k=10$ gives a very high precision result.
The functions $\tilde{\mathcal{Y}}_k$, $\tilde{\mathcal{M}}_k$, $\tilde{\mathcal{D}}_k$ and $\tilde{\mathcal{Q}}_k$ can be written as (see CNSN): 
\begin{align} 
\label{eqn:YMDQ_def} 
    \tilde{\mathcal{Y}}_k =& \sum_{j=1}^{2k+2} a_j^{k} \mathcal{\hat{O}}_{\tilde{x}}^j \tilde{n},\nonumber\\
    \tilde{\mathcal{M}}_k =& \sum_{j=1}^{2k+2} \frac{a_j^{k}}{6}\left[j(j+2)\mathcal{\hat{O}}_{\tilde{x}}^j+(2j+3)\mathcal{\hat{O}}_{\tilde{x}}^{j+1}+\mathcal{\hat{O}}_{\tilde{x}}^{j+2} \right]\tilde{n},\nonumber\\
    \tilde{\mathcal{D}}_k =& \sum_{j=0}^{2k+2} d_j^{k} \left[j \mathcal{\hat{O}}_{\tilde{x}}^j+\mathcal{\hat{O}}_{\tilde{x}}^{j+1} \right] \tilde{n},\nonumber\\
    \tilde{\mathcal{Q}}_k =& \sum_{j=0}^{2k+2} \frac{q_j^{k}}{3} \left[j(j-1)\mathcal{\hat{O}}_{\tilde{x}}^j+2j \mathcal{\hat{O}}_{\tilde{x}}^{j+1}+\mathcal{\hat{O}}_{\tilde{x}}^{j+2} \right] \tilde{n}.
\end{align}
For convenience we defined the differntial operator $\mathcal{\hat{O}}^k_{\tilde{x}}=\tilde{x}^k\partial^k_{\tilde{x}}$.
The coefficients $a_j^k$, $d_j^k$ and $q_j^k$ can be found tabulated in CNSN and are directly related to the moments, $I^k_{\ell m}$. They are part of {\tt SZpack}\footnote{{\tt SZpack} is a computational package used for the fast and accurate computation of the SZ effect from hot, moving clusters.} and available up to 10$^{\rm th}$ order in $\thetae$. 
Multiple scattering effects were neglected \citep{Chluba2014mSZI, Chluba2014mSZII}

In Eq.~\eqref{eqn:Dn_asym}, $\tilde{\mathcal{Y}}_k$ describes the thermal SZ effect (tSZ) with relativistic temperature corrections, also called the relativistic SZ effect (rSZ); $\tilde{\mathcal{D}}_k$ and $\tilde{\mathcal{Q}}_k$ are the first and second order (dipole and quadrupole) kinematic corrections, with their higher order temperature terms; $\tilde{\mathcal{M}}_k$ contains the kinematic correction to the monopole signal, and its higher order temperature corrections. 
No temperature-independent terms arise as Thomson scattering ($\propto \tau \betac^2$) of the radiation monopole does not modify the field.
The combination of dipole, quadrupole and monopole correction, together give the kinematic SZ effect (kSZ) to second order in $\betac$. Their temperature corrections are, in general, also referred to as relativistic SZ effects. All of these are examined in more detail in CNSN.

\subsubsection{Final signal in the CMB rest frame}
\label{sec:MathsCMBframe}
The details of converting between frames are explored in detail in CNSN. However, here we will summarise and expand on these results. To transform the distortion into the CMB rest frame, we must perform a Lorentz transformation on our variables. In particular, as we have already defined the relative speed $\betac$, we find
\begin{equation} \label{eqn:var_cmb_frame} \begin{split}
    &\tilde{x} = \gammac x(1-\betac \muc), \\
    &\muct = \frac{\muc-\betac}{1-\betac \muc}.
\end{split} \end{equation}
Due to the invariance of the photon occupation number, all one has to do is to use these expressions in Eq.~\eqref{eqn:Dn_asym} when evaluating the distortion.
Here, $\tilde{x}$ and $\muct$ are defined in the cluster frame as before, and $x$ and $\muc$ are defined in the CMB frame; $\betac$ (and accordingly $\gammac=1/\sqrt{1-\smash{\betac^2}}$) is identical in both frames. In principle, one might also consider the modifications of $\thetae$, $\tau$ and $y$ between the two frames; however, we will always use these quantities defined in the {\it cluster frame}, since this is the frame in which these values have a clear physical meaning. Further justifications of this can be found in CNSN.

Alternatively, instead of transforming the variables before calculating the results, we can substitute the relations Eq.~\eqref{eqn:var_cmb_frame} into Eqs.~\eqref{eqn:Dn_asym} and \eqref{eqn:YMDQ_def}, and then expand up to second order in $\betac\ll 1$. After some simplification we can then write [compare Eq.~(25) of CNSN]
\begin{equation} \label{eqn:Dn_asym_cmb} \begin{split}
    \Delta n \simeq &\tau\thetae\sum_{k=0}^\infty \thetae^k \mathcal{Y}_k + \tau\betac^2\left(\left[x\partial_x+\frac{1}{3}x^2\partial_x^2\right]n+\thetae\sum_{k=0}^\infty \thetae^k \mathcal{M}_k\right) \\
    &\quad +\tau \muc \betac \left(-x\partial_x n+\thetae\sum_{k=0}^\infty \thetae^k \mathcal{D}_k\right)\\
    &\qquad +\tau P_2(\muc)\betac^2\left(\frac{11}{30} x^2\partial_x^2 n + \thetae \sum_{k=0}^\infty \thetae^k \mathcal{Q}_k\right).
\end{split} \end{equation}
Here, we have not only modified the functions to $\mathcal{Y}_k$, $\mathcal{M}_k$, $\mathcal{D}_k$ and $\mathcal{Q}_k$, but we have also modified the leading order terms to the monopole correction and quadrupole terms, caused by boosting and aberration effects.
The transformed distortion functions become\footnote{These functions are also used explicitly in {\tt SZpack} but the expressions were not given in this form in CNSN}
\begin{align} 
\label{eqn:YMDQ_def_cmb}
    \mathcal{Y}_k &= \tilde{\cal{Y}}_k
    \nonumber\\
    \mathcal{M}_k &= \sum_{j=1}^{2k+2} \frac{a_j^{k}-d_j^k}{3}\left[j(j+2)\mathcal{\hat{O}}_{x}^j+(2j+3)\mathcal{\hat{O}}_{x}^{j+1}+\mathcal{\hat{O}}_{x}^{j+2} \right]n,\nonumber\\
    \mathcal{D}_k &= \sum_{j=0}^{2k+2} (d_j^{k}-a_j^{k}) \left[j \mathcal{\hat{O}}_{x}^j+\mathcal{\hat{O}}_{x}^{j+1} \right] n,\\ \nonumber
    \mathcal{Q}_k &= \sum_{j=0}^{2k+2} \frac{q_j^{k}-2d_j^{k}+a_j^{k}}{3} \left[j(j-1)\mathcal{\hat{O}}_{x}^j+2j \mathcal{\hat{O}}_{x}^{j+1}+\mathcal{\hat{O}}_{x}^{j+2} \right] n.
\end{align}
We note that $a_0^k=0$, since this is not explicitly mentioned in CNSN. We also note that when doing these conversions directly, many of the corrections come from expressing $n(\tilde{x})$ in terms of an expansion in $\betac$ around $n(x)$. With these expressions we can easily obtain the radio SZ signal in the cluster and CMB rest frames. 
We highlight that at second order in $\betac$, the expressions in the CMB rest frame {\it do not agree} with previous works, as was clarified in CNSN. Luckily, corrections at the level of $\betac^2$ usually remain subdominant. However, recent works on using terms $\propto \betac\thetae$ to learn about large-scale structure \citep[e.g.,][]{Coulton2020} may also be affected, as subtle differences in these terms arise (see CNSN).


\subsection{The radio SZ signal}
For the normal SZ effect as caused by the CMB, we use the blackbody occupation of $n_\cmb = 1/(\expf{x}-1)$. 
The radio background has been estimated in \citet{Fixsen2011}. These results were used in \citet{Holder2021}, where the radio photon density was expressed as
\begin{equation} \label{eqn:RadioBackground}
    n_\rR = \frac{A_\rR\,f(z)}{x_0}\left(\frac{x}{x_0}\right)^{-\alpha}.
\end{equation}
Here, $\alpha = 3.59\pm0.04$ is the radio spectral index, $A_\rR=8.84\pm0.77$ is the normalization at $z=0$, $x_0= 5.46\times10^{-3}$ and $f(z)$ is the fraction of the radio excess that is present at a given redshift to account for any change in the radio background over time. If the radio background is fully cosmological and formed at much higher redshifts than the clusters, we would expect $f(z)=f(z=0)=1$. This formulation also assumes that there is no variation in $\alpha$ with redshift. We note that $x$ is redshift independent, so as with the typical SZ signal, there will be no redshift evolution of the signal beyond that caused by the redshift evolution of the background i.e., $f(z)$ itself.

The derivatives of the CMB density are complicated, but can be written in closed form (see Appendix~A2 of CNSN) as
\begin{equation}
    x^k \partial_x^k n_\cmb = \left(\frac{-x}{1-\expf{-x}}\right)^{k}\sum_{m=0}^{k-1} \left\langle \begin{matrix} k\\m\end{matrix}\right\rangle \expf{-m x}\,n_\cmb,
    \label{eq:SZ_case}
\end{equation}
where the coefficients $\left<...\right>$ denote the Eulerian Numbers. The latter determine the number of permutations of the numbers 1 to $m$ in which exactly $k$ elements are greater than the previous element.
The derivatives of the Planckian exhibit very strong frequency dependence in the Wien-tail ($x\gg 1$). This implies that relativistic corrections can in principle be used to measure the cluster temperature and velocity.

In contrast, for a radio-like background with $n_\rR = A x^{-\alpha}$, we find
\begin{equation} \label{eqn:gamma_derivs}
    x^k \partial_x^k n_\rR = (-1)^k \frac{(\alpha+k-1)!}{(\alpha-1)!}\,n_\rR.
\end{equation}
This expression implies that the relativistic corrections to the radio SZ signal do not alter the spectral shape of the radio signal, and they in total combine merely to a change in the amplitude of the signal. This means the cluster temperature and velocity cannot be independently determined using only the radio contribution.

It is worth noticing that, since at low frequencies the CMB background can be approximated as a power-law with $\alpha=1$ (i.e., $n_\cmb\approx 1/x$), the rSZ effects are equally well behaved at $x\ll 1$ and also do not show any new spectral shape. Indeed from Eq.~\eqref{eq:SZ_case}
\begin{equation}
    x^k \partial_x^k n_\cmb \approx \left(-1\right)^{k}\sum_{m=0}^{k-1} \left\langle \begin{matrix} k\\m\end{matrix}\right\rangle \,n_\cmb=
    \left(-1\right)^{k} k!\, n_\cmb,
    \label{eq:SZ_case_small_x}
\end{equation}
where we used the identity $\sum_{m=0}^{k-1} \left\langle ...\right\rangle=k!$, which directly follows from the combinatorial definition of the Eulerian numbers. This result agrees with the one obtained from Eq.~\eqref{eqn:gamma_derivs} for $\alpha=1$.

\vspace{-3mm}
\subsubsection{Expression inside cluster frame}
Using the expressions from above, we can simplify Eq.~\eqref{eqn:YMDQ_def} to
\begin{equation} \label{eqn:YMDQ_gamma} \begin{split}
    \tilde{\mathcal{Y}}_k^\rR &= \tilde{n}_\rR \sum_{j=1}^{2k+2} (-1)^j \frac{(\alpha+j-1)!}{(\alpha-1)!} a_j^{k},\\
    \tilde{\mathcal{M}}_k^\rR &= \tilde{n}_\rR \sum_{j=1}^{2k+2} \frac{(-1)^j}{6} \frac{(\alpha+j-1)!}{(\alpha-1)!} a_j^{k} \alpha(\alpha-2),\\
    \tilde{\mathcal{D}}_k^\rR &= \tilde{n}_\rR \sum_{j=0}^{2k+2} (-1)^j \frac{(\alpha+j-1)!}{(\alpha-1)!} d_j^{k} (-\alpha),\\
    \tilde{\mathcal{Q}}_k^\rR &= \tilde{n}_\rR \sum_{j=0}^{2k+2} \frac{(-1)^j}{3} \frac{(\alpha+j-1)!}{(\alpha-1)!} q_j^{k} \alpha(\alpha+1).
\end{split} \end{equation}
These are the expressions for the radio background with general $\alpha$ that can be used in Eq.~\eqref{eqn:Dn_asym}. For $\alpha=1$ they equally apply to the standard SZ effect at low frequencies.
The first few terms of these can then be written as 
\begin{equation} \label{eqn:YMDQ_gamma_explicit}\begin{split}
    \tilde{\mathcal{Y}}_0^\rR &= \tilde{n}_\rR \alpha(\alpha-3),\\
    \tilde{\mathcal{Y}}_1^\rR &= \frac{\tilde{n}_\rR}{10} \alpha(\alpha-3)(7\alpha^2-21\alpha-3),\\
    \tilde{\mathcal{Y}}_2^\rR &= \frac{\tilde{n}_\rR}{120} \alpha(\alpha-3)(44\alpha^4-264\alpha^3+284\alpha^2+336\alpha-31),\\
    \tilde{\mathcal{M}}_0^\rR &= \frac{\tilde{n}_\rR}{6} \alpha^2(\alpha-3)(\alpha-2),\\
    \tilde{\mathcal{D}}_0^\rR &= \frac{\tilde{n}_\rR}{5} 2\alpha(\alpha^2-3\alpha+1),\\
    \tilde{\mathcal{D}}_1^\rR &= \frac{\tilde{n}_\rR}{5} \alpha(2\alpha^4-12\alpha^3+16\alpha^2+6\alpha+1),\\
    \tilde{\mathcal{Q}}_0^\rR &= \frac{\tilde{n}_\rR}{30} \alpha(\alpha+1)(\alpha^2-3\alpha-6).
\end{split} \end{equation}
For the monopole and quadrupole terms, we only give the leading order corrections $\propto \tau\betac^2\thetae$. Note also that $\tilde{\mathcal{Y}}_0^\rR$ is the main term of the radio SZ effect \citep{Holder2021}.
Inserting $\alpha = 3.59$ and $\alpha = 1$, to second order in temperature and cluster velocity,
\begin{subequations}
\begin{align} 
\label{eqn:app_rad_num}
    \frac{\Delta \tilde{n}_\rR}{\tilde{n}_\rR} &\approx  2.12y\big[1+1.18\thetae-0.59\thetae^2\big]+2.01\betac^2\,y
    \\ \nonumber
    &\!\!\!\!\!\!\!\!\!\!\!\!
    +3.59\muct\betac\big[\tau+y(1.25+1.15\thetae)\big]
    -4.94P_2(\muct)\betac^2\big[\tau+0.43 y\big],
\\
\label{eqn:app_cmb_num} 
    \frac{\Delta \tilde{n}_\cmb}{\tilde{n}_\cmb} &\approx  -2y\big[1-1.7\thetae+3.08\thetae^2\big]+0.34\betac^2\,y
    \\ \nonumber
    &\,\,\,
    +\muct\betac\big[\tau+y(-0.4+2.6 \thetae)\big]
    -0.6 P_2(\muct)\betac^2\big[\tau+0.89y\big],
\end{align} 
\end{subequations}
respectively. These expressions are quite accurate as we will see below, but higher order terms can be easily added in {\tt SZpack}. 
We emphasize that for the individual contributions no new spectral shape is created by relativistic corrections. 
However, since the radio and CMB SZ effects have a slightly different dependence on the temperature and cluster velocity, the spectrum of the total signal is modified.

\vspace{-5mm}
\subsubsection{Expression inside CMB frame}
Much like for the cluster frame, in the CMB frame using $n_\rR$ we can simplify the terms of Eq.~\eqref{eqn:YMDQ_def_cmb} for use with Eq~\eqref{eqn:Dn_asym_cmb} to
\begin{equation} \label{eqn:YMDQ_gamma_cmb} \begin{split}
    \mathcal{M}_k^\rR &= n_\rR \sum_{j=1}^{2k+2} \frac{(-1)^j}{3} \frac{(\alpha+j-1)!}{(\alpha-1)!} \left(a_j^{k}-d_j^{k}\right) \alpha(\alpha-2),\\
    \mathcal{D}_k^\rR &= n_\rR \sum_{j=0}^{2k+2} (-1)^j \frac{(\alpha+j-1)!}{(\alpha-1)!} \left(d_j^{k}-a_j^{k}\right) (-\alpha),\\
    \mathcal{Q}_k^\rR &= n_\rR \sum_{j=0}^{2k+2} \frac{(-1)^j}{3} \frac{(\alpha+j-1)!}{(\alpha-1)!} \left(q_j^{k}-2d_j^{k}+a_j^{k}\right) \alpha(\alpha+1).
\end{split} \end{equation}
We did not repeat the $\mathcal{Y}_k^\rR$, which are identical to $\tilde{\mathcal{Y}}_k^\rR$ in Eq.~\eqref{eqn:YMDQ_gamma_explicit}.
The first few terms in the CMB frame can now be written as
\begin{equation} \begin{split}
    \mathcal{M}_0^\rR &= \frac{n_\rR}{5} 7\alpha^2(\alpha-3)(\alpha-2),\\
    \mathcal{D}_0^\rR &= \frac{n_\rR}{5} \alpha(7\alpha^2-21\alpha+2),\\
    \mathcal{D}_1^\rR &= \frac{n_\rR}{10} \alpha(11\alpha^4-66\alpha^3+92\alpha^2+21\alpha+2),\\
    \mathcal{Q}_0^\rR &= \frac{n_\rR}{10} \alpha(\alpha+1)(19\alpha^2-57\alpha+2).
\end{split} \end{equation}
As we will see below, with these expressions we reproduce the kinematic and relativistic temperature corrections to the radio SZ (and standard SZ) effect to high precision.

Again inserting $\alpha = 3.59$ and $\alpha = 1$, we find
\begin{subequations}
\begin{align} 
\label{eqn:app_rad_num_cmb} 
    \frac{\Delta n_\rR}{n_\rR} &\approx  2.12y\big[1+1.18\thetae-0.59\thetae^2\big]+9.08\betac^2\big[\tau+0.62y\big]
    \\ \nonumber
    &\!\!\!\!\!\!\!\!\!\!\!\!
    +3.59\muc\betac\big[\tau+y(3.37+3.65\thetae)\big]
    +6.04P_2(\muc)\betac^2\big[\tau+3.84 y\big],
    \\
\label{eqn:app_cmb_num_cmb}
    \frac{\Delta n_\cmb}{n_\cmb} &\approx  -2y\big[1-1.7\thetae+3.08\thetae^2\big]
    +1.67\betac^2\big[\tau+0.56y\big]
    \\ \nonumber
    &
    +\muc\betac\big[\tau+y(-2.4+6 \thetae)\big]
    +0.73 P_2(\muc)\betac^2\big[\tau-3.27y\big].
\end{align}
\end{subequations}
Our expression for the temperature corrections to the standard thermal SZ effect agrees exactly with Eq.~(37) in \citet{Nozawa1998SZ}, even if here we give only two orders in $\thetae$, which is already highly accurate. Also the leading order kinematic term $\simeq \muc \betac \tau$ agrees; however, all other terms depart as explained in CNSN.

It is important to stress that the temperature terms appearing in the form $y\thetae^k$ should be interpreted as $y$-weighted temperature moments, which for non-isothermal clusters can differ from mass-weighted or X-ray measured temperatures \citep{Chluba2012moments, Lee2020}. Similarly, all other terms are caused by various temperature and velocity moments that should be treated carefully. However, for the current discussion, we neglect this complication.

We can also immediately see that the temperature corrections to the radio SZ signal are smaller than those for the CMB signal in this low frequency regime. To first order, we can see that since the CMB signal is negative in the region, the two signals combine destructively when they are added. This leads to a source-shadow appearance of the cluster in the scattered light \citep{Holder2021}. However, the temperature corrections themselves to both the CMB and radio SZ signal are in the same direction, and as a result magnify the measured relativistic corrections measured in either signal individually. We further observe that the kinematic corrections are conversely larger for the radio SZ signal than for the CMB signal.

\subsection{Anisotropies in the radio background} 
\label{sec:anis_radio_maths}
In the absence of kinematic effects, we can compute the scattering effect for arbitrary photon anisotropies as in \citet{Chluba2012}. Neglecting stimulated scattering terms, we can express the distortion to first order in temperature as
\begin{equation} \label{eq:photon_anisotropy} \begin{split}
    \Delta n \approx \tau&\left(n_0+\frac{n_2}{10}-n\right)+y \left[4x\partial_x + x^2\partial_x^2\right]\bar{n}\\ &+y\left(-\frac{2}{5}n_1-\frac{3}{5}n_2+\frac{6}{35}n_3\right), \\
    \bar{n}=n_0&-\frac{2}{5}n_1+\frac{1}{10}n_2-\frac{3}{70}n_3.
\end{split} \end{equation}
Here $n_{\ell}\equiv \sum_{m}n_{\ell m} Y_{\ell m}(\pmb{\hat{\gamma}} \cdot \pmb{\hat{z}})$, where $\pmb{\hat{z}}$ defines the coordinate system for the multipole expansion. Thus $n_0$, $n_1$, $n_2$ and $n_3$ refer to the monopole, dipole, quadrupole and octupole components, and  $n$ refers to the combined signal. 
We can combine this explicitly by expanding and taking $\alpha_\ell$ as the spectral index for each $n_\ell$ to find
\begin{equation} \label{eq:photon_anisotropy_expanded} \begin{split}
    \Delta n \approx n_0 &\Big[y\,\alpha_0(\alpha_0-3)\Big] -\frac{1}{5} n_1\left[5\tau+2y(\alpha_1^2-3\alpha_1+1)\right]\\
    &-\frac{1}{10}n_2\left[9\tau-y(\alpha_2^2-3\alpha_2-6)\right]\\
    &\qquad-\frac{1}{70}n_3\left[70\tau+3y(\alpha_3^2-3\alpha_3-4)\right].
\end{split} \end{equation}
It is important to note that these are equivalent to the expression we find in Eq.~\eqref{eqn:Dn_asym} to first order in temperature if one defines the multipoles as those derived from a kinetic boost from an isotropic CMB frame photon into the cluster frame [cf. Eq.~(11) in CNSN].
That is, we can immediately observe the similarities between the terms in Eq.~\eqref{eq:photon_anisotropy_expanded} and those for $\tilde{\mathcal{Y}}_0^\rR$, $\tilde{\mathcal{D}}_0^\rR$ and $\tilde{\mathcal{Q}}_0^\rR$ found in Eq.~\eqref{eqn:YMDQ_gamma_explicit}.
In principle, it would therefore be possible to extend the expressions for the monopole, dipole and quadrupole scattering to higher order in temperature, as we have for the general SZ effect. 
In this case, no extra terms to the monopole scattering would appear [e.g., $\mathcal{\tilde{M}}=0$]. Also, terms describing octupole scattering [equivalent to $\mathcal{O}(\betac^3)$ in Eq.~\eqref{eqn:Dn_asym} which have been dropped] would still need to be derived.
However, below we focus on a low-temperature expansion to more easily understand the effects of various anisotropies. We also stress that to first order in temperature {\it only} the first three anisotropies could ever contribute to the SZ signal.

For a simplistic model, where the radio background has, for instance, a dipole with the same spectral shape as the background itself, say $n_1 = f_1 n_\rR$, with $f_1 < 1$, then the effect on the {\it radio} signal would be the same as the first order (in $\betac$) kinematic corrections with $\betac \muct \alpha \leftrightarrow f_1$. However, a kinematic effect would also create a correction to the incoming CMB photons, which an intrinsic dipole anisotropy in the radio background would evidently not  -- accordingly these two signals would in principle be distinguishable. Furthermore, if any radio background anisotropies have a different spectral shape to the radio background itself, we would see more complex contributions to the final signal.
Below, we will consider instructive examples to illustrate the effects.

\vspace{-3mm}
\subsection{Effect of the observer motion} 
In Sect.~\ref{sec:MathsCMBframe}, we discussed the effects of switching between the cluster frame and CMB frame and the effects on the measured SZ signal. In actuality, of course, observations are made in neither. It is well known that there is a significant observed dipole component within the CMB due to the earth's motion relative to the CMB frame. This can be well parameterised by $\betao=1.241\times10^{-3}(1\pm0.2\%)$ \citep{Fixsen1996, Fixsen2002}. We can then define $\muo$ as the cosine of the angle between this dipole direction, and the observed cluster (measured in the {\it observer frame}).

In a similar way to moving to the CMB frame from the cluster frame, this dipole motion will generate both a dipole correction and higher order terms to the observed SZ signal. This has already been well documented \citep{Chluba2005b} for its effect on the conventional SZ signal, and the effects on radio SZ signal are similar, but worth repeating here for completeness.

From a mathematical perspective, this is all most easily expressed by simply transforming our variables such that (see CNSN)
\begin{equation} \begin{split}
    &x = \gammao x_{\rm o}(1+\betao\muo) \\
    &\tilde{x} = \gammac x(1-\betac\muc) = \gammac\gammao x_{\rm o}(1-\betac\muc)(1+\betao\muo) \\
    &\mu = \frac{\muo+\betao}{1+\betao\muo}
\end{split} \end{equation}
where $x_{\rm o}$ is the dimensionless frequency as measured in the observer frame and $\gammao = \sqrt{1-\betao^2}$ is the associated Lorentz factor; $\mu$ is the angle between the observer dipole and the observed cluster as measured in the {\it CMB frame} and is included here only for completeness.

It is clear that for both the radio SZ and normal SZ signals, the effect of this frame conversion is a modulation of the observed frequency. If this was to be fully expanded as in Sect.~\ref{sec:MathsCMBframe} for the CMB frame, it would introduce dipole and quadrupole terms in $\betao$ alongside cross terms in $\betac \betao$. However, it is easier to simply replace variables in the CMB frame expression using the above relation to obtain the transformed result. Since we know the amplitude and direction of the dipole, we can easily take this effect into account. We thus assume this has been taken care of in any analysis.

\subsection{Additional small effects} 
In our discussion, we omitted multiple scattering effects. These can lead to additional small corrections to the SZ signals \citep{Chluba2014mSZI, Chluba2014mSZII}. In addition, the scattering of the intra-cluster radio light, e.g., from radio sources within the cluster requires a more detailed modelling.
Internal motions of the cluster, e.g., due to cluster rotation \citep{Chluba2002, Cooray2002, Diego2003} could affect the radio SZ signal in a similar manner as for the normal SZ effect.
Similarly, moving lens effects \citep{Birkinshaw1983, Molnar2003} will appear. 
Finally, like with the usual kinematic SZ effect \citep{Sunyaev1980} and scattering of the CMB quadrupole \citep{Kamionkowski1997b, Sazonov1999}, the scattered radio light will be linearly polarized. 
However, a modelling all these smaller effects is beyond the scope of this work.

\vspace{-3mm}
\section{Results}
\label{sec:results}
In this section, we consider how the radio and CMB signal combine to give the observable radio SZ signal; the kinematic and relativistic contributions modify the signal and affect the location of the radio SZ null. In Sect.~\ref{sec:background_var}, we explore how variations in the radio background change the results and in Sect.~\ref{sec:anis_radio_results}, we show the contributions of anisotropies in the radio background. Here we will always give our results in the {\it CMB frame}.

\begin{figure}
    \includegraphics[width=0.95\linewidth]{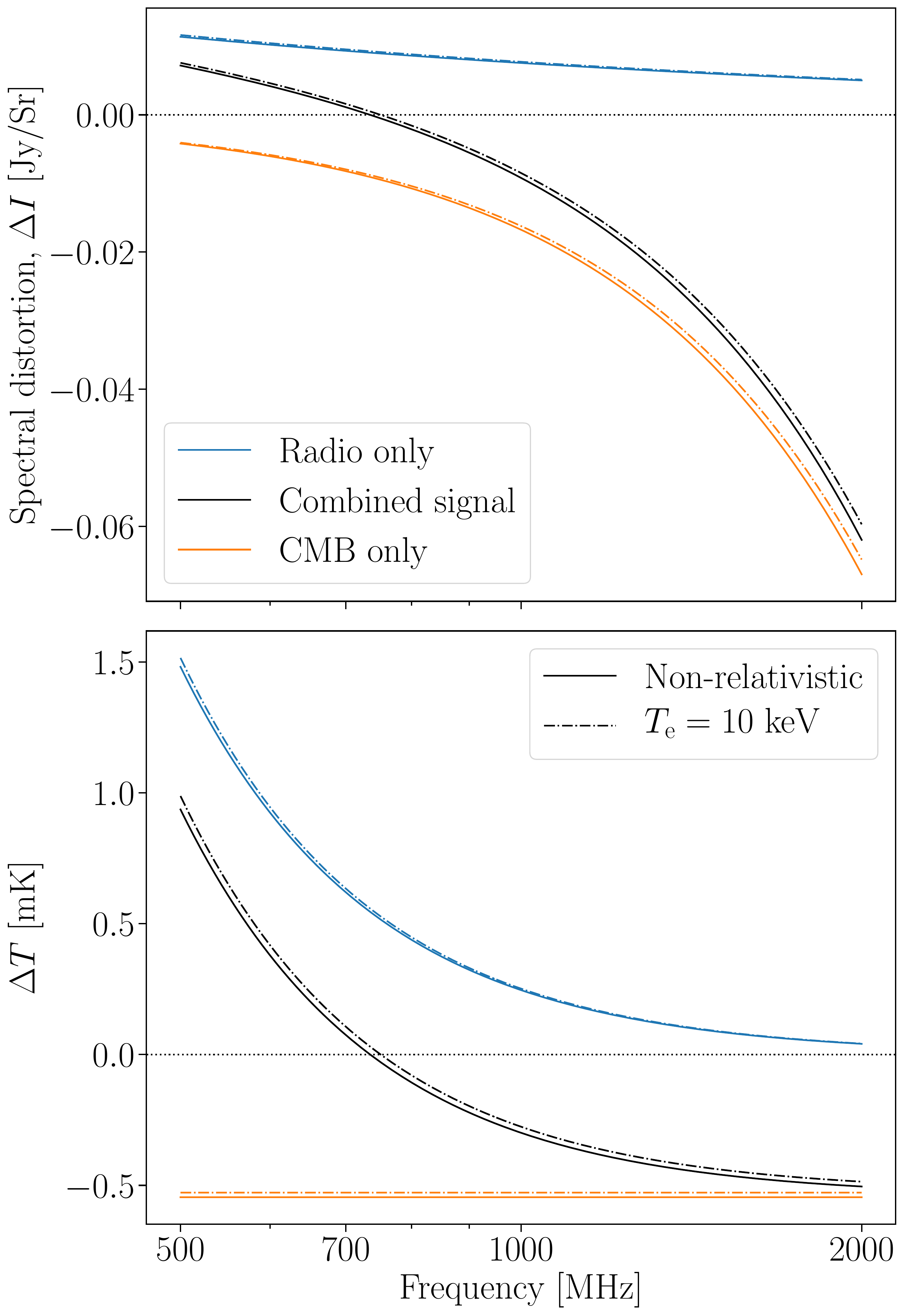}
    \\
    \caption{Contributions of the non-relativistic CMB SZ signal, radio SZ signal and combination of the two at low frequencies, alongside their relativistic corrections at $T_{\rm e} = 10$~keV. In the top panel we see the change in intensity relative to the background signals, the bottom panel shows the effective change in temperature measured at each frequency. We have set $y=10^{-4}$ and used $n_\rR$ with $f(z)=1$. Note this is a logarithmic plot in frequency.}
    \label{fig:non_rel}
\end{figure}

\vspace{-3mm}
\subsection{Main dependence of total signal and degeneracies}
From Eq.~\eqref{eqn:app_rad_num_cmb} [or in general the combination of Eqs~\eqref{eqn:gamma_derivs}, \eqref{eqn:Dn_asym} and \eqref{eqn:YMDQ_def}] it is evident that the radio background causes a scattering signal that is always proportional to the radio background itself. i.e., 
\begin{equation} \begin{split}
    \Delta n_\rR &= \tau n_\rR(\nu, z, \alpha) F_\rR(\thetae, \betac, \muc, \alpha)\\&= \tau A_\rR f(z) F_\rR(\thetae, \betac, \muc, \alpha)\,g(\nu, \alpha),
\end{split} \end{equation}
where $F_\rR$ now contains all of the kinematic and relativistic terms and can in general be thought of as a scattering amplitude. Here we have also defined $g(\nu,\alpha)=(x/x_0)^{-\alpha}\,x_0^{-1}$.

We note that the radio background is always positive and thus does not form an inherent null from scattering. In contrast, the normal CMB SZ signal has an inherent null at around 217~GHz, the cross-over frequency from photons being scattered out from lower frequencies into higher frequencies. However, the combination of scattering of the radio background and CMB leads to a new null at around 735.5~MHz, as discussed in \cite{Holder2021}. That is, we can write
\begin{equation} \begin{split}
    \Delta n_\rR + \Delta n_\cmb \approx \tau &\Big[A_\rR f(z) F_\rR(\thetae, \betac, \muc, \alpha)\,g(\nu,\alpha) \\&\qquad+ F_\cmb(\thetae, \betac, \muc)\,n_\cmb(\nu)\Big],
\end{split} \end{equation}
where we note that $F_\cmb(\thetae, \betac, \muc)\approx F_\rR(\thetae, \betac, \muc, 1)$, and we note that $F_\cmb$ is negative in this regime. This can all be seen easily in Fig.~\ref{fig:non_rel}, where the scattering from each the radio background and the CMB are shown individually alongside the combined signal.

We can now quantify any changes to the observed signal in a heuristic manner. The kinematic and relativistic effects cause changes to $F_\rR$ and $F_\cmb$, which to first order enhance the observed change to the signal. e.g., in Fig.~\ref{fig:non_rel} we show the relativistic corrections to each signal for a cluster temperature of 10~keV. We see that the relativistic effect on both the CMB and radio scattering increase the signal. We can understand this physically as the relativistic corrections implying there is more up-scattering for the radio SZ signal, while for the CMB distortion this results in fewer photons being up-scattered out of this region.
We will quantify the relativistic corrections more precisely below, but at low frequencies, the effects are typically a few percent. These are comparable, if a little smaller, to the standard SZ effect.

The kinematic corrections combine in a similar way -- a cluster heading towards the observer (i.e., $\muc>0$) leads to more up-scattering, shifting the null to higher frequencies, while a cluster heading away shows the opposite effect. Changes to the amplitude of the radio background -- either due to the uncertainty in measurements of $A_\rR$, from $f(z)$ or from some inherent angular dependence to the signal, i.e., some $f(\pmb{\hat{\gamma}}, z)$, also leads to a re-weighting of the radio component relative to the CMB component.

Here we can immediately observe a degeneracy within the signal. For a given $\alpha$, measurements of the total radio SZ signal can only ever give the relative amplitudes of the CMB and radio components -- i.e., from the shape of the combined signal. However, the more details of the clusters or the radio background are known, the more information can be obtained from the signal. For instance, if $\tau$ (or $y$) is known, the amplitude of the signal itself becomes a second independent measurement from the radio SZ signal. It should also be noted that variations in $\alpha$, either within the measurement of the radio background itself, or due to anisotropies, would in principle be distinguishable from the other variations within the signal. Furthermore, since $F_\rR$ and $F_\cmb$ are not independent quantities and together give a complex dependence on $T_{\rm e}$, $\betac$ and $\muc$, more information may be gained from them than initially supposed. However, this requires detailed forecasts and high precision measurements.

\vspace{-2mm}
\subsection{Relativistic and kinematic corrections}
\begin{figure}
    \includegraphics[width=0.95\linewidth]{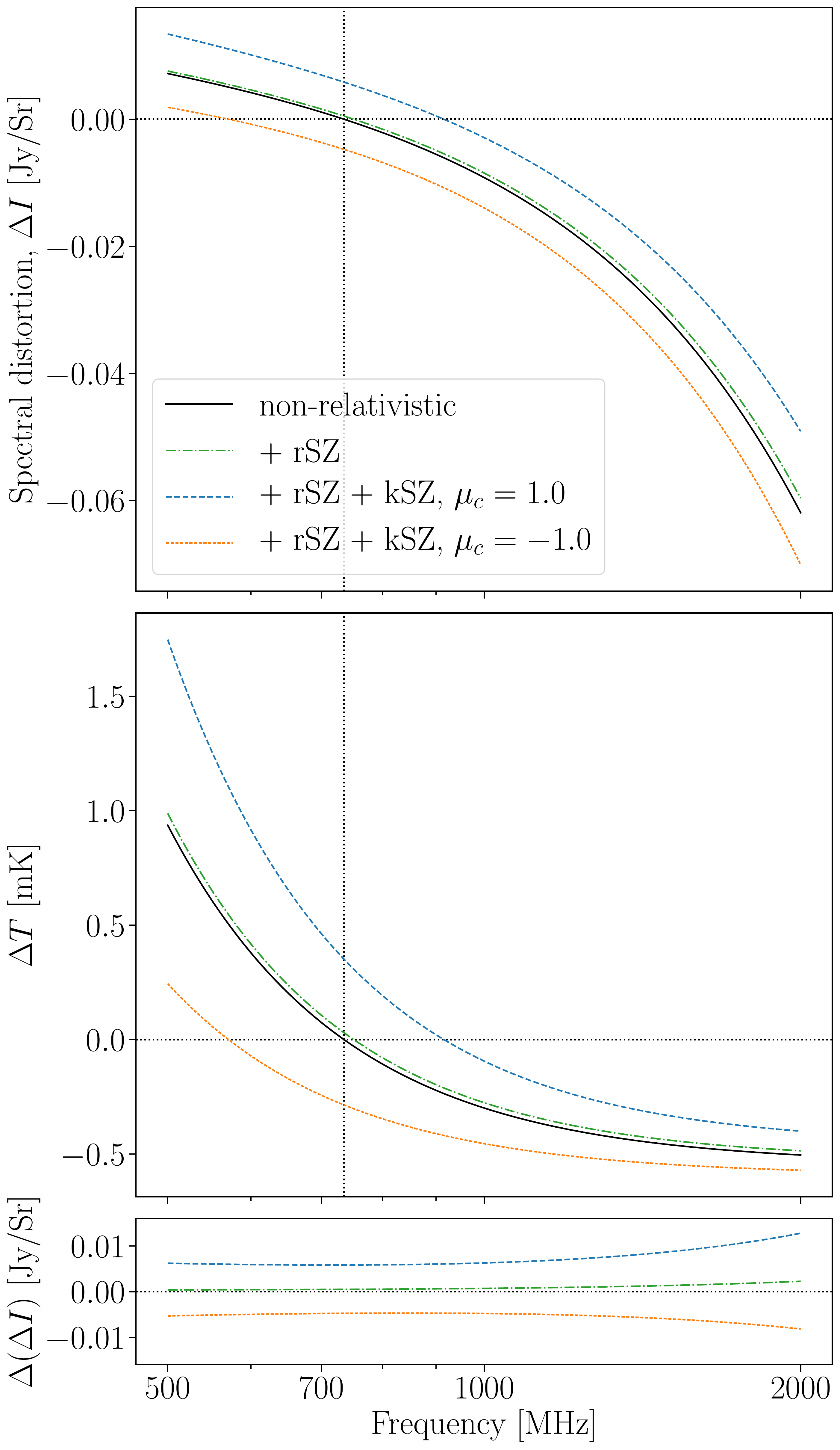}
    \caption{Illustration of the corrections to the combined CMB and radio SZ signals due to relativistic (rSZ) and kinematic (kSZ) effects. The rSZ corrections come from $T_{\rm e} = 10$~keV and the kSZ effects are for $\betac=0.005$, with $\muc = \pm 1.0$. In the top panel we see the change in intensity relative to the background signals, the middle panel shows the effective change in temperature measured at each frequency. The bottom panel here shows the difference between the stationary non-relativistic signal and the signals incorporating rSZ and kSZ effects. We have set $y=10^{-4}$ and used $n_\rR$ with $f(z)=1$. Note this is a logarithmic plot in frequency.}
    \label{fig:relkin}
\end{figure}

In Figure \ref{fig:relkin}, we show the non-relativistic combined signal \citep[as described in][]{Holder2021} alongside the signal including relativistic and kinematic corrections. Here we have shown an energetic cluster in terms of the kinematic corrections (with $\betac=0.005$ and $\muc=\pm1.0$). The relativistic corrections come from a cluster temperature, $T_{\rm e} = 10$~keV, a temperature typical of the most massive clusters in the Universe \citep{Arnaud2005, Lee2020}. We have calculated all of these distortions with {\tt SZpack}.\footnote{The modifications for the specific code here are available on request.}
We assume here that the radio background is comoving with the CMB -- or alternatively that it is isotropic in the CMB rest frame. That is, that it is meaningful to use the same value of $\betac$ for both the CMB induced signal and the radio SZ signal. 

The first aspect to note is that while these corrections roughly maintain the shape of the distortion in these regions they lead to shifts in frequency. A second point is that the shapes are nonetheless distorted, as can be seen in the third panel of Figure 
\ref{fig:relkin}. The kinematic effects are generally larger than the relativistic effects, as in the normal SZ effect. In comparison to the standard kSZ effect, the kinematic radio SZ effect is $\simeq 3.6$ times larger due to the steepness of the radio background. 

We can also note that the kinematic effects are not reversible with respect to $\muc$. That is, for example, the $\muc=1.0$ signal contributes to more up-scattering of the photons, than the $\muc=-1.0$ signal contributes to down-scattering compared to the thermal SZ signal. This is a small effect in the figure shown, but proportional to $\betac^2$. This can be seen numerically from Eq.~\eqref{eqn:Dn_asym}, where we see that the signal has components proportional to $\betac^2$ and $\betac^2 P_2(\muc)$. For a similar reason, we find that the changes to the signal are non-linear under changes to $\muc$. Also, the $\betac^2 \tau$ terms in the radio SZ effect are amplified by a factor of $\simeq 7$, which causes an larger asymmetry when flipping the direction of $\betac$ in comparison to the standard kSZ effect. However, direct observations of these small terms will be difficult in the near future, but modelling of these effects will lead to improved accuracy in any  measurements.

\subsection{Corrections to the null}
We can illustrate the corrections if we consider the adjustments to the null in the signal. We define the null frequency, $\nu_{\rm N}$ to be the frequency at which there is no distortion to the original signal (i.e., $\Delta I(\nu_{\rm N}) = 0$~Jy/Sr and $\Delta T(\nu_{\rm N}) = 0$~$\mu$K). For the normal SZ effect this is often referred to as the crossover frequency. 
\begin{figure}
    \includegraphics[width=0.95\linewidth]{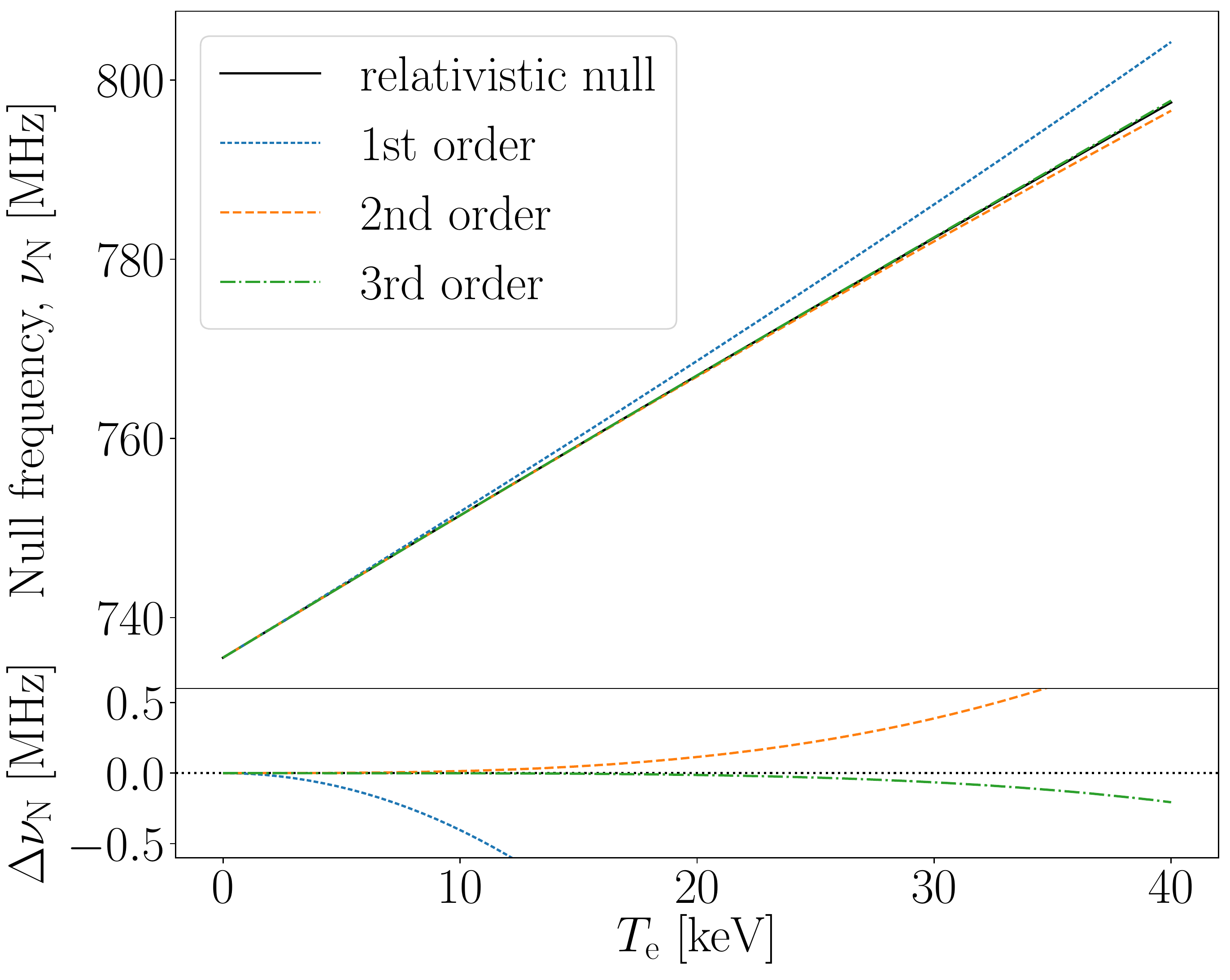}
    \caption{A plot showing the relativistic corrections to radio null. In black we can see this value accurately calculated, while the other lines show the result of the asymptotic expansion to 1st, 2nd and 3rd order. The second panel shows the difference between these approximations and the true value. We can see here that a low-order expansion remains very accurate even to very high electron temperatures.}
    \label{fig:rel_null}
\end{figure}

In Figure \ref{fig:rel_null}, we show the variation of the radio null under changes in $T_{\rm e}$. Kinematic terms are omitted. 
Firstly, we see that the changes are small, and secondly the variations are almost linear with cluster temperature. The figure also illustrates the convergence of the asymptotic expansion, by showing the calculation of the null from considering increasing numbers of terms. In Table \ref{tab:relativistic_nulls}, we specifically calculate the null for a selection of temperatures.
By interpolating these points, one can obtain accurate predictions for the null. However, at $T_{\rm e} \lesssim 15\,{\rm keV}$, the expressions in Eq.~\eqref{eqn:app_rad_num_cmb} and \eqref{eqn:app_cmb_num_cmb}
should suffice. Higher precision results can be obtained with {\tt SZpack}.

\begin{table}
\caption{The null of the radio+cmb SZ signal for various temperatures. Kinematic terms are excluded. We note that since the relativistic null is nearly linear with temperature (see Fig.~\ref{fig:rel_null}), the null can be well modelled as an interpolation between these points.} \centering
\begin{tabular}{l |@{}c |@{}c |@{}c |@{}c |@{}c |@{}c}
    \hline
    $T_{\rm e}$ [keV] & $0$ & $5$ & $10$ & $20$ & $30$ & $40$\\
    \hline
    $\nu_{\rm N}$ [MHz] & 735.5 & 743.5 & 751.4 & 767.0 & 782.4 & 797.5 \\
    $\Delta\nu_{\rm N}/\nu_{\rm N}$ 
    & $--$ & $1.08\%$ & $2.16\%$ & $4.28\%$ & $6.37\%$ & $8.43\%$\\
    \hline
\end{tabular}
\label{tab:relativistic_nulls}
\end{table}

The null frequency, $\nu_{\rm N}$ increases with temperature, reflecting the general shift upwards in both the CMB and radio contributions. However the effect of increasing the temperature, decreases with temperature, e.g., the shift in null from 0 to 10~keV is slightly greater than that between 30 and 40~keV. The response of the null to changes in temperature is, however, well behaved and relatively small compared to the frequencies considered.

In the standard SZ scenario, the crossover frequency can be parameterised under changes in temperature as \citep{Itoh98},
\begin{equation}
    \nu_{\rm N, CMB} = 217.5\, \left[1+1.1674\thetae-0.8533\thetae^2\right]\,{\rm GHz}.
    \label{eq:SZ_null}
\end{equation}
The equivalent general function for the radio SZ effect is more complicated -- but allowing for changes in $\alpha$, $A_\rR$ and $f(z)$ we find
\begin{equation} \begin{split}
    \nu_{\rm N} &\approx \nu_0\left(\frac{2}{A_\rR f(z) \alpha(\alpha-3)}\right)^\frac{1}{1-\alpha} \bigg[1+\frac{7}{10}(\alpha-2)\thetae
    \\&\qquad\quad+\frac{1}{600}(73\alpha^3-218\alpha^2-730\alpha+1748)\thetae^2+ O(\thetae^3)\bigg]
\end{split} \end{equation}
where $\nu_0 = 310$~MHz is the real frequency associated with the dimensionless frequency $x_0$. For general $\alpha$ this is complicated to assess, but for $\alpha=3.59$ we can simplify this as
\begin{equation}
    \nu_{\rm N} \approx 735.5\left(\frac{A_\rR f(z)}{8.84}\right)^{0.386} 
    \big[1+1.113\thetae-0.5079\thetae^2\big]\,{\rm MHz}.
\end{equation}
Comparing to Eq.~\eqref{eq:SZ_null} shows that the relative effect of temperature corrections on the radio null is similar to that for the normal tSZ effect, which enter at the few percent level.

A similar full consideration for the kinematic effect is more complex -- however, for $\muc\betac\ll \thetae$ the lowest order kinematic correction can be expressed as 
\begin{equation} \begin{split}
    \nu_{\rm N} &\approx \nu_0\left(\frac{2}{A_\rR f(z) \alpha(\alpha-3)}\right)^\frac{1}{1-\alpha} 
    \bigg[1+\frac{1}{2(\alpha-3)}\frac{\muc\betac}{\thetae}\bigg].
\end{split} \end{equation}
Again inserting numbers we then find
\begin{equation}
    \nu_{\rm N} \approx 735.5\left(\frac{A_\rR f(z)}{8.84}\right)^{0.386} \left[1+0.847\,\frac{\muc\betac}{\thetae}\right]\,{\rm MHz}.
\end{equation}
This expression immediately shows that kinetic terms can have a relatively large effect on the radio null.

\begin{figure}
    \includegraphics[width=\linewidth]{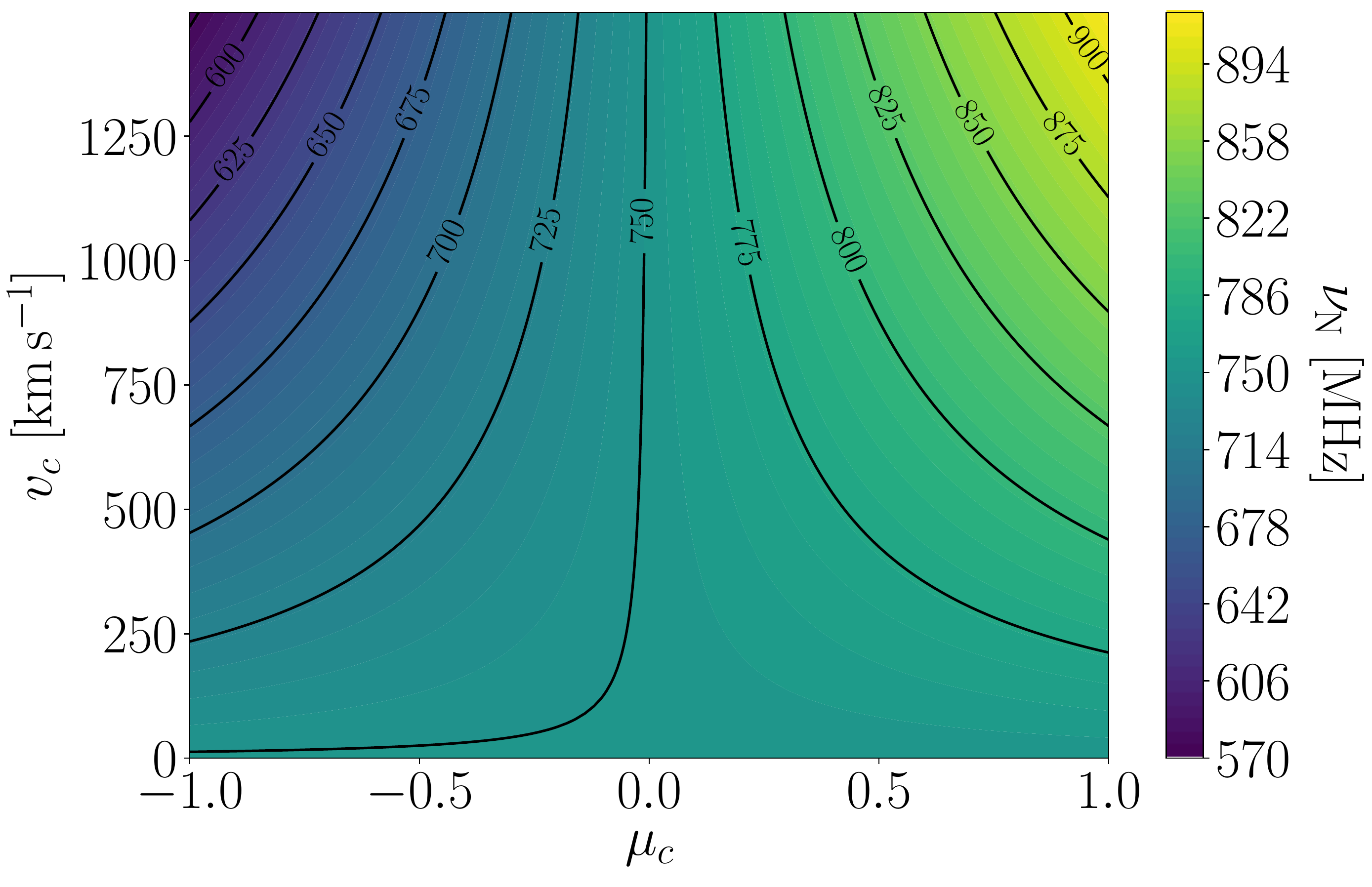}
    \\
    \caption{A plot showing the kinematic corrections to radio null. Here we are calculating everything with relativistic corrections of $T_{\rm}=10$~keV. We can see the response of the null frequency, $\nu_{\rm N}$, to changes in viewing angle, $\muc$, and cluster velocity, $\varv_{\rm c} = c \betac$. A slight asymmetry around $\muc=0$ is visible, which is due to terms $\tau\betac^2$.}
    \label{fig:rkin_null}
\end{figure}
An indicative plot for the kinematic corrections is shown in Fig.~\ref{fig:rkin_null}, which shows the response of the null to $\betac$ and $\muc$ at $T_{\rm e} = 10$~keV.
As expected, even relatively small peculiar motion can induce significant shifts in the radio null. One can observe a slight asymmetry in $\muc$ that is more easily seen at higher values of $\betac$. The kinematic corrections also to first order vary about the relativistically corrected null -- i.e., when $\muc = 0$, the null lies at $\simeq751.4$~MHz as expected from Table \ref{tab:relativistic_nulls} -- although at high $\betac$ the monopole corrections will have more of an effect in shifting this to higher frequencies. 

It is worth noting that, as with the CMB SZ null, there is an anticorrelation between the cluster temperature and the shift in the null. That is, at lower temperatures the kinematic effects become more important than at higher temperatures for a fixed $\betac$. This can be understood in a number of ways -- numerically, we can directly see that since $y\simeq \tau \thetae$, even the non-relativistic tSZ signal has a dependence proportional to $\thetae$, while the non-relativisitic kSZ signal has none. Hence, decreasing the cluster temperature, reduces the impact of the tSZ signal, thus effectively increasing the impact of the kSZ signal. Alternatively, we can understand this energetically, in that, for a fixed velocity cluster, (i.e., fixed $\betac$), reducing the temperature of the cluster will lead to a higher proportion of the electron energy being contained in the kinematic component. Accordingly the kinematic effect will increase proportionally with lower temperatures.

\subsection{Radio background variations}
\label{sec:background_var}
The radio background is still comparatively poorly understood, despite the measurements that have been taken \citep{Fixsen2011, Seiffert2011, Singal2018, Dowell2018}. As such the uncertainty in the background is far larger than that of the CMB. Not only is it currently unknown how the radio background varies with redshift, there may also be large-scale spatial fluctuations within the radio background. The radio SZ signal could theoretically then be used as a measure to determine details of the radio background itself alongside potentially bounding its inherent variations.

\begin{figure}
    \includegraphics[width=\linewidth]{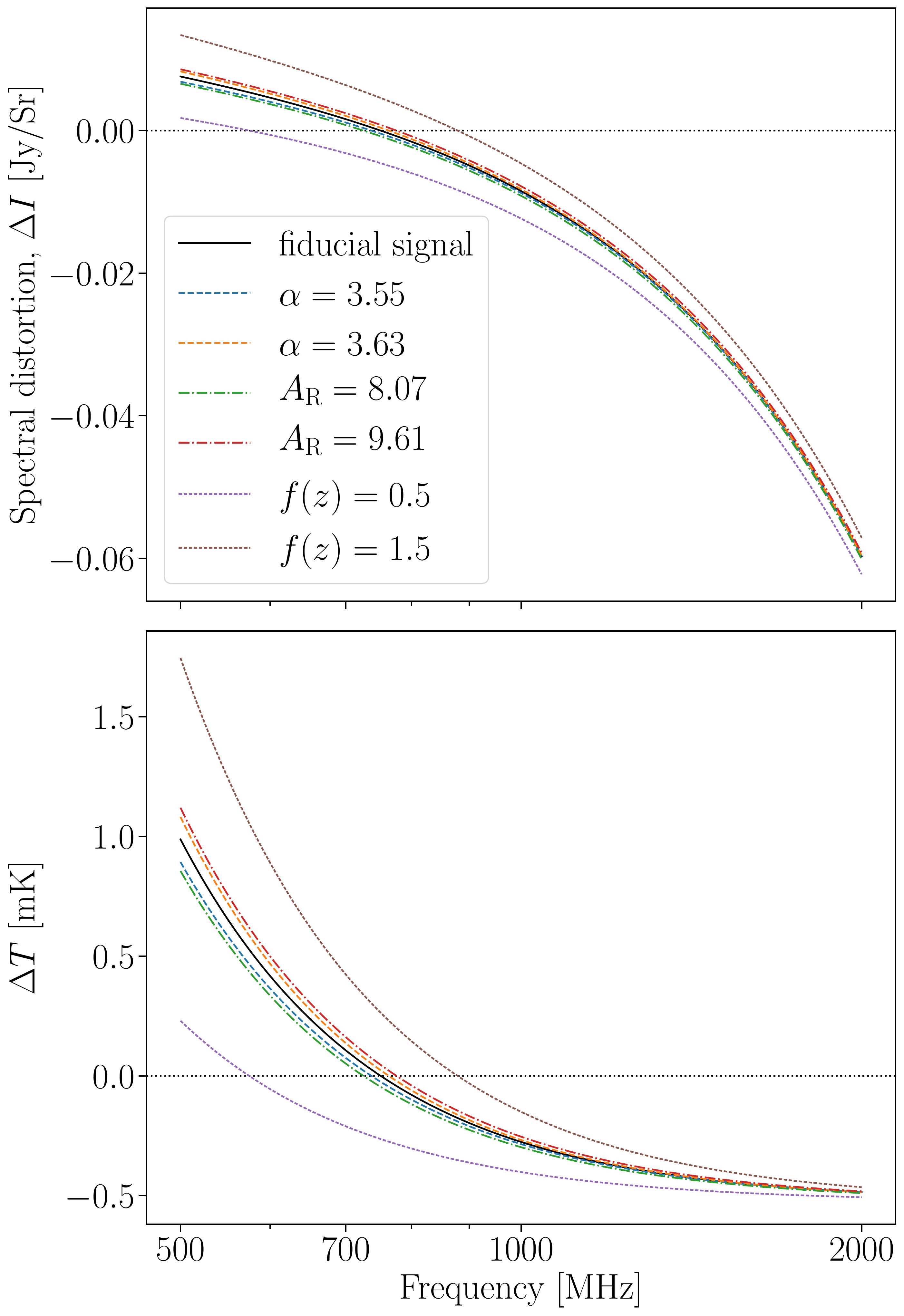}
    \\
    \caption{Variations in the radio SZ signal (with $\betac =0$ and $T_{\rm e}=10$~keV) under modifications to the radio background distribution. The dashed lines show variations due to the error on $\alpha$, the dashed-dotted lines show the variations of the measured error in the amplitude as written in Eq.~\eqref{eqn:RadioBackground}, and the dotted lines show variations according to a significant variation in $f(z)$. We have set $y=10^{-4}$.}
    \label{fig:signal_variations}
\end{figure}

In Figure \ref{fig:signal_variations}, we illustrate the effects of the $1\sigma$ errors given for the radio background as expressed in Eq.~\eqref{eqn:RadioBackground} in both the normalization and spectral index. We can see immediately that these changes are of a similar scale to the relativistic corrections, but small compared to any major changes from $f(z)$, for which we illustrate the two cases, $f(z)=0.5$ and $f(z)=1.5$. A value $f(z)<1$ is consistent with the simple picture that the radio background is isotropic but slowly builds up with redshift, while $f(z)>1$ suggests that a local overdensity around the cluster might be present. 

The shown examples all change the shape of the signal (as would be expected). This means that these variations are theoretically separable from the relativistic corrections caused by the cluster temperature. 
However, the normalization $A_\rR$ of $n_\rR$, $f(z)$ and the effective relativistic corrections are all degenerate, and only their product can be constrained.
Existing priors, either from standard SZ observations or from theory using scaling relations, could thus provide additional important leverage when aiming to use radio SZ measurements to learn about $f(z)$ and $A_\rR$. 
The spectral index $\alpha$ on the other hand can in principle be separated from the other effects by measuring the precise spectral shape of the radio SZ signal.

The shift in the null from varying the spectral index, $\alpha$, is around $\simeq \pm14$~MHz, while the measured uncertainty in $A_\rR$ leads to a slightly larger effect of around $\simeq \pm25$~MHz variation to the null. We note that the scattered radio SZ signal need not have the same spectral index as the `off-cluster' radio flux. This is because the scattered signal indeed probes the local radio background at the clusters location, while the `off-cluster' sky flux constraints the superposition of all radio emission along the line of sight. The value of $\alpha$ should therefore be directly measured using the scattered signal.

Overall, assuming that we can precisely measure the temperature and peculiar velocity of a cluster, a high precision measurement of the radio SZ signal can allow a determination of $\alpha$ from the signal shape, and the product $A_\rR f(z)$ from the location of the signal null. The exact details of the current observational feasibility are beyond this paper, but it is worth noting that any large changes over redshift could be more easily detectable. Large changes in $f(z)$ may indicate the age of the radio background or more insight into its creation. It is also worth noting the possibility of the variation of $\alpha$ with redshift, $\alpha\rightarrow\alpha(z)$, which again could theoretically be measured  with a large enough sample size and high enough measurement precision.

Other variations could come if $\alpha$ had any $x$ dependence -- for example, if the radio background exhibited significant curvature beyond the power law. A superposition of different power laws (i.e., a mix of  different values for $\alpha$) would lead to some curvature inherently \citep{Chluba2017foregrounds}; however, there is no evidence for this within the ARCADE measurements. This indicates that there is little variation of $\alpha$ with redshift. A detailed analysis of curvature within the power law background is however beyond the scope of this paper.

\subsection{Anisotropy in the radio background} \label{sec:anis_radio_results}
We can also consider the effects of anisotropy in the radio background on the observed signal. Overall, this can be thought of as $A_\rR f(z)\rightarrow A_\rR f(\pmb{\hat{\gamma}}, z)$. 
In Sect.~\ref{sec:anis_radio_maths} we detailed the mathematical formalism and noted that if the anisotropic components have the same spectral index, the only effect will be to change the amplitude of the radio component of the SZ signal. That is, as discussed before, this would shift the null. Hence, if it were possible to bound the average $f(z)=\int f(\pmb{\hat{\gamma}}, z)\,{\rm d} \hat{\gamma}$, $T_{\rm e}$ and the amplitude of the radio background itself, a precise measurement of the null would give some bounds on the level of anisotropy in the radio background itself. If however, the higher multipole components have a different spectral index, we would see changes to the spectral shape as we would have more components adding to the distortion (as opposed to just the monopole radio $n_\rR$ and CMB components).
This would increase the number of parameters and pose high demands on the observations.

\begin{figure}
    \includegraphics[width=0.95\linewidth]{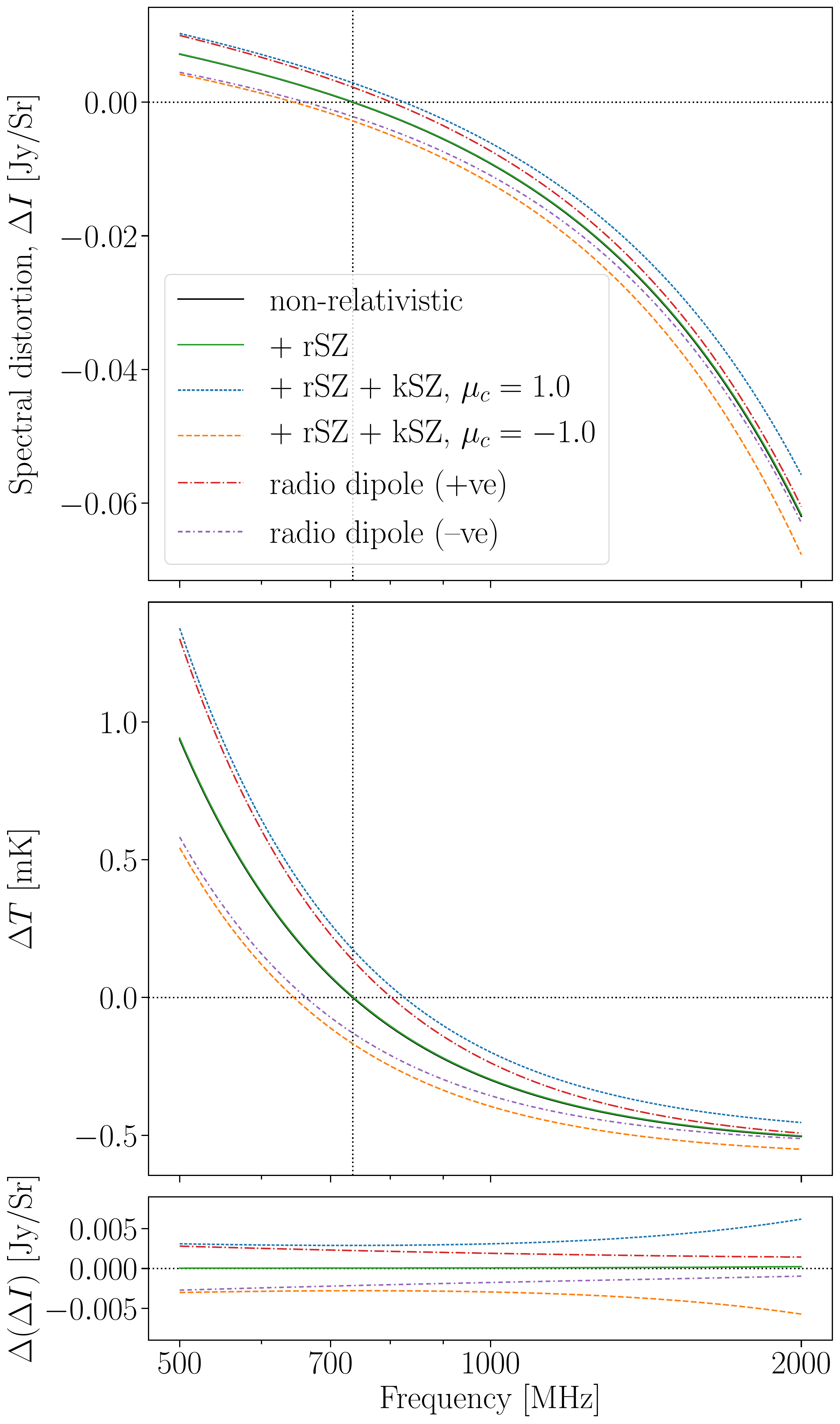}
    \\
    \caption{A plot showing the differences between a dipole in the radio background and the kinematic corrections. With the exception of the non-relativistic line, all these signals take relativistic corrections according to $T_{\rm e}=1$~keV. Note, this is a significantly smaller $T_{\rm e}$ than our previous figures. The radio dipole comes from a signal $n_1=\pm0.001 n_\rR$, while the kSZ effects use $\betac=2.79\times10^{-4} = 0.001/\alpha$, with $\muc = \pm 1.0$. In the top panel we see the change in intensity relative to the background signals, the middle panel shows the effective change in temperature measured at each frequency. The bottom panel here shows the difference between the stationary, non-relativistic signal and the other signals. We have set $y=10^{-4}$ and used $n_\rR$ with $f(z)=1$. Note this is a logarithmic plot in frequency.}
    \label{fig:ani_dipole}
\end{figure}

In Figure~\ref{fig:ani_dipole}, we show the effects of a dipole in the radio background and its differences to the kinematic effect. Here we use a substantially cooler cluster of 1~keV, both to increase the accuracy of the first order temperature expansion we use in Eq.~\eqref{eq:photon_anisotropy} and to maximise the size of the distortions, so they can be more clearly observed. Here we set the radio background dipole to be 0.1\% of the radio background as a whole (i.e., $n_1=\pm0.001 n_\rR$) and we also plot the kinematic effects with the `equivalent' magnitude, i.e., $\betac\muc=0.001/\alpha$. We note that the dipole in the CMB (induced by the relative motion of the observer frame and the CMB frame) is around $1.24\times10^{-3}$ (i.e., $\betao$). So here we are assuming a dipole of comparative size but caused by the matter distribution.

We can immediately see that these two effects have very different effects on the shape of the observed signal. At low frequencies, the kinematic corrections to the CMB component are small, so the radio dipole and kinematic effect cause similar distortions. However, at higher frequencies, the CMB part of the kinematic effect is a larger component and the radio dipole and kinematic corrections diverge. These two effects are therefore observationally distinct. However, more work may be necessary to determine the possible radio anisotropies and to fully determine the radio background itself, before it would be possible to ascertain the full details of the radio background anisotropies.
A similar discussion can follow for the radio quadrupole and octupole. Since these are not expected to be connected to the CMB equivalents, the spectral effects should be distinguishable and possibly larger than for the scattering of the CMB temperature anisotropies.

\section{Discussion}
\label{sec:discussion}
It is worth taking a moment to consider the potential of different observational strategies in determining any of the discussed features in the radio SZ signal. A detailed forecast is {\it far} outside the scope of the of this work, but here we will discuss the differences between stacking approaches and individual cluster observations when it comes to the specifics of the radio SZ signal itself. Many of the ideas can be directly adapted from normal SZ observations.

If we have a catalog of clusters, each will sit in a unique region of space -- they will have a distribution of temperatures, redshifts and peculiar velocities (both $\betac$ and $\muc$), leading to variations in $F_\rR(\thetae, \betac, \muc, \alpha)$. There is also the possibility that the background radio signal itself depends on redshift and position of the cluster [i.e., variations in $f(\pmb{\hat{\gamma}}, z)$]. If there is any spatial or redshift dependence in $\alpha$ these will also modify our observed signal between clusters. 

It is clear that the more we know about each cluster in our sample, (1) the better we can model the expected radio SZ signal and (2) the more certainty we can have on the cause of any measured variations in the radio background. For nearby clusters, for example, we may already know $y$, $\betac$, $\muc$ and $T_{\rm e}$ through the standard SZ signal measurements, or through a combination of analogous measurements, e.g., X-ray observations. In such a situation, variations in the observed signal must all come from variations in $A_\rR f(z)$, in $\alpha$ (which would both change the shape of the radio contribution $g(\nu, \alpha)$ and the scaling through $F_\rR$) or through some more complicated anisotropies within the radio background. Conversely, of course, if our measurements of the radio background are significantly more constrained than those for the cluster's measurements, then we may be able to determine the parameters of clusters through the radio SZ measurement itself -- as is conventional within the typical SZ experiments.

If information on individual clusters is less precise, we could resort to stacking analyses made over large samples of clusters. 
For the normal SZ effect, this has been attempted using {\it Planck} data to constrain the rSZ effect \citep{Hurier2017rSZ, Erler2017}. In \cite{Erler2017}, they used to 772 clusters to find an averaged $\langle y\rangle=(1.24\pm0.04)\times10^{-4}$ with an averaged SZ temperature of $k_{\rm B}T_{\rm SZ}=4.4^{+2.1}_{-2.0}$~keV. Since the radio SZ signal also depends on $y$ and $k_{\rm B}T_{\rm SZ}$, we could expect to find a signal of a similar magnitude under the stacking procedure.

Applied to radio SZ measurements, we would be able to deduce averages of many of the parameters. For the simplest stacking procedure, we would then be averaging over temperature, peculiar velocity and redshift of the catalog. Since the peculiar velocities can be assumed to be uncorrelated, almost all kinematic corrections would be averaged out, that is we would perform $\langle F_\rR(\thetae, \betac, \muc, \alpha) \rangle_{\muc}$. This would leave kinematic terms only in the monopole corrections at $O(\betac^2)$, which are generally very small. As such, any measured stacked relativistic corrections would indicate the sample-averaged cluster temperatures, as well as the spatially- and redshift-averaged radio background.
Again, prior knowledge on the temperature distribution of the clusters can be used to refine our understanding of the deduced radio background parameters. 

If stacking procedures were then performed on subsets of the catalogue, it would then be possible to bound variations in $f(\pmb{\hat{\gamma}},z)$ within the radio background itself -- i.e., (1) by spatially selecting clusters, it would theoretically be possible to directly constrain anisotropies in the radio background, or (2) by using priors on the redshift location of clusters, to determine the redshift evolution (e.g., redshift binning). This is very similar in spirit to using clusters as distributed observers of for the remote CMB quadrupole (and dipole) anisotropy \citep{Deutsch2018}.
As for the normal SZ signal, by using priors on the orientation of the clusters' peculiar velocities \citep{Hand2012,Cayuso2021}, it may be possible to learn about the averaged cluster $\langle\betac\rangle$.
In contrast to the standard SZ effect, for the radio SZ effect anisotropies are not necessarily limited to being very small. This could greatly increase the observability of the radio SZ effects. For example, if there are small-scale fluctuations in the radio background then the `on-cluster' and `off-cluster' background may be distinct, e.g., if the cluster itself were a significant source of the radio background. 

It is also worth noting however, that a full sky signal would likely prove very difficult to detect for the radio SZ signal. In \cite{Hill2015}, they predict that for the typical SZ signal, a whole sky average would result in $\langle y\rangle\simeq1.7\times10^{-6}$, and a $y$-weighted temperature of $\langle k_{\rm B}T_{\rm SZ}\rangle \simeq 1.3$~keV.\footnote{The conversion to $y$-weighted temperature was used in \citet{abitbol_pixie}.} While for the typical SZ signal this may be measured with future high precision spectroscopy instruments such as PIXIE \citep{Kogut2011PIXIE}, or Voyage 2050 \citep{Voyage2050}, no such instruments are planned which may reach this level of precision in the radio SZ regime over the full sky.

\vspace{-3mm}
\section{Conclusion}
In this paper we derive the corrections to the radio SZ signal arising from kinematic terms, relativistic electron temperature corrections and possible anisotropies in the radio background. The main goal was to give all the expressions that are required to describes these effects and to illustrate the main similarities and differences to the equivalent effects on the standard SZ effect.
Our discussion is merely meant to stimulate the ideas about next steps related to observability of these effects and how one could potentially go about it. A detailed forecast is beyond the scope of this work.

We find that, similarly to the standard SZ signal, the kinematic effects cause the largest effects for the radio SZ signal, with these effects being $\gtrsim 2.5$ times larger in the radio SZ than for the standard CMB SZ signal. These effects must always be accounted for in single cluster measurements, and can be easily modelled. For stacking calculations, however, the kinematic effects will be greatly reduced, as the orientations of the clusters peculiar velocities will `cancel' most of their contributions in large stacking considerations. 

The relativistic temperature corrections are marginally smaller in the radio SZ signal than in the standard SZ signal (around a few percent over reasonable cluster temperatures). These corrections are also around the same size as those from the current errors in the measurement of the radio background from \citep{Fixsen2011}, indicating that to get competitive bounds on the radio background itself, would require a similar precision of detection.
However, as such it also possible to use the radio SZ signal in concert with the normal SZ signal to obtain higher confidence prior measures of cluster properties themselves. 
SZ clusters are now routinely observed with e.g., Planck \citep{Planck2016ymap}, ACT \citep{ACT2020Choi} and SPT \citep{Bleem2015}, and
with future SZ samples obtained with SPT-3G \citep{Benson2014SPIE.9153E..1PB}, the Simons Observatory \citep{SOWP2018}, CMB-S4 \citep{Abazajian2016S4SB} and CCAT-prime \citep{CCATp2018}, many of the required observables could become available.

Our analysis shows that, if there are anisotropies in the radio background, or if the background has redshift evolution, these may well be far larger effects (see Fig.~\ref{fig:ani_dipole}). We find that any radio background anisotropies lead to contributions which scale similar to kinematic effects. Both of these could be observationally explored with a stacking analyses on a large enough cluster sample (see Sect.~\ref{sec:discussion}). The radio SZ signal thus provides a unique tool to probe the nature of the radio background in determining whether it is truly `cosmological' (in which case we would expect little to no redshift evolution).

While a forecast of the details of measuring this signal, are beyond this paper, this work provides a detailed description of the contributing factors to the radio SZ signal itself, in the hope that future forecasts might be eased. Furthermore, with the advent of more large-scale, high precision experiments such as the SKA \citep{Dewdney2009, Bacon2020} and MeerKAT \citep{Jonas2016Meerkat}, there is the potential that these measurements may not be quite as futuristic as they might seem. Another obvious way forward is through cross-correlation studies, which are also frequently used in normal SZ observations \citep{Hand2012}.

While here we focused our discussion on the scattering of the radio background, this merely provides a mathematically simple example of examining the effects of scattering on background photon sources beyond only the CMB. Thus, much of the work presented here can be extended to other backgrounds, e.g., the CIB, the 21cm line, X-rays emitted by clusters themselves \citep[e.g.,][]{Cooray2006, Grebenev2020}.
All our expressions should be directly applicable to CIB scattering at frequencies below the CIB maximum, which essentially behaves like a power-law with $\alpha_{\rm CIB}\simeq 0.14$, where we quoted the average CIB spectrum as used by the {\it Planck} collaboration \citep{Planck2013components, Planck2016GNILC}.
We look forward to extending our formalism to these cases in the future, noting that in particular scattering of the CIB could already have an important effect on current SZ analysis targeting the relativistic SZ \citep[e.g.,][]{Erler2017}.

\vspace{-3mm}
{\small
\section*{Acknowledgements}
EL was supported by the Royal Society on grant No.~RGF/EA/180053.
JC was supported by the Royal Society as a Royal Society University Research Fellow at the University of Manchester, UK (No.~URF/R/191023).
This work was also supported by the ERC Consolidator Grant {\it CMBSPEC} (No.~725456) as part of the European Union's Horizon 2020 research and innovation program.
GH received support from CIfAR and Brand \& Monica Fortner.
}

\section*{Data Availability}
The code used in this article will be available on request to the authors. A full extension of the {\tt SZpack} code is planned in the near future, including the code used here. The original form of {\tt SZpack} can be found at www.chluba.de/SZpack.

{\small
\bibliographystyle{mnras}
\bibliography{Radio_rSZ} 
}

\appendix


\bsp	
\label{lastpage}
\end{document}